# Enhanced charge-transfer character in the monoclinic phase of Mott-insulator LaVO$_3$ thin film

Anupam Jana, Rajamani Raghunathan, Ritu Rawat, R. J. Choudhary* and D. M. Phase

UGC DAE Consortium for Scientific Research, Indore- 452 001, India

**Abstract:**

Electronic structure of pulsed laser deposited epitaxial LaVO$_3$ (LVO) thin film grown on LaAlO$_3$ (001) (LAO) substrate, has been studied at room temperature and at 130 K, which is below the structural (141 K) and magnetic transition temperatures (143 K) of single crystal LVO. Significant modification in spectral intensity, largely around the V 3$d$ - O 2$p$ hybridized region, is observed in valence band spectrum (VBS) of LVO film at 130 K. Resonant photoemission study at 130 K confirms the presence of charge transfer screened 3$d^n\underline{L}$ ($\underline{L}$: hole in the O 2$p$) final state along with the dominant 3$d^{n-1}$ final state at 1.5 eV binding energy in the valence band. On the contrary, in the room temperature VBS dominant V 3$d$ sates with only 3$d^{n-1}$ kind final state is accentuated. To understand this difference, density functional theory (DFT) calculations are employed. The changes in crystal structure from room temperature orthorhombic (O-LVO) to low temperature monoclinic (M-LVO) symmetry leads to remarkable change in the electronic structure around the Fermi level, including transition from direct to indirect nature of band gap. Our calculations also confirm an enhanced O 2$p$ character in the valence band edge of M-LVO that is hybridized with V 3$d$. These results are further corroborated with octahedral distortions associated with the structural transition.

*Corresponding Author: ram@csr.res.in,

**Introduction:**

Transition metal oxides (TMOs) constitute a very attractive class of materials owing to the manifestation of a wide variety of exotic phenomena such as Mott Metal-insulator transition (MIT), colossal magnetoresistance (CMR), ferroelectricity and superconductivity [1-5]. Even though decades of extensive research have been made by studying the late TMOs (involving Cu, Ni, Fe and Mn), these phenomena are yet to be completely understood. The microscopic origin of these properties are driven by the delicate interplay between the on-site Coulomb repulsion ($U_{dd}$), Hund's direct exchange (J), charge transfer ($\Delta$), crystal field splitting ($\Delta_{CF}$) and metal-ligand hybridization ($T$) energies [6]. Even though the effect of electron correlations on electronic structure and magnetic properties has been well studied in late 3$d$ TMOs, early 3$d$-TMOs such as vanadium oxides, with less than half filled $d$-orbitals have received less attention [7-10]. Recently, interest on vanadates, in particular $LaVO_3$ (LVO), has gained momentum after the reports of emergence of two dimensional insulating or metallic conductivity for the $n$-type and $p$-type $LaVO_3/SrTiO_3$ interface [11-12] as well as the room temperature ferromagnetism in superlattices systems involving $LaVO_3$ and $SrVO_3$ [13]. These intriguing properties have introduced a new pathway to manipulate new states of the materials in thin film forms by maneuvering the substrate induced strain [14]. Thus, LVO films grown on $LaAlO_3$ substrate is known to show Mott-Hubbard type insulating character at room temperature [15], unlike its bulk counterpart which is considered to be intermediate between Mott-Hubbard and charge transfer insulator characters [16-18]. LVO is known to exhibit a structural transition from orthorhombic to monoclinic below 141 K and magnetic transition from paramagnetic to antiferromagnetic below 143 K [19,20]. Iso-electronic $V_2O_3$ also undergoes a Mott-Hubbard type first order metal to insulator transition (MIT) at around 155 K from a high temperature paramagnetic metal to low temperature antiferromagnetic insulating phase, accompanied by a structural transition from rhombohedral to monoclinic phase [1,21]. No such MIT was observed across structural and magnetic transition of bulk LVO [22]. However, a drastic change in resistance, known to be order-disorder transition, was observed in $LaVO_3/SrVO_3$ superlattices across the structural transition of LVO [23]. The ordering of in-equivalent V sites in the monoclinic structure of LVO was suggested to be responsible for this order-disorder transition.

It is known that GdFeO$_3$-type distortion in VO$_6$ octahedra leads to a stabilization of orthorhombic structure of LVO at room temperature [19]. Along this line, a recent study made on LaVO$_3$ film grown on SrTiO$_3$ substrate demonstrated that under compressive stress, orthorhombic structure of LaVO$_3$ undergoes further distortion in which the VO$_6$ octahedra rotate to accommodate the substrate induced strain [24]. Such distortions in VO$_6$ octahedra has immense effect on the bond topology and the metal-ligand hybridization. This can consequently change the energy landscape resulting in unique electronic and magnetic properties of the LVO films, that are not seen in bulk samples [11,13,15,25]. Though there are a few reports available on the electronic structure of orthorhombic LVO [12,26,27], the same has not been explored in the monoclinic phase. Understanding the modification in the electronic structure of LVO going from orthorhombic to monoclinic structure is thereby very crucial to comprehend the electrical and magnetic properties of the hetero-structures comprising of LVO [23,28,29]. In the present experimental cum theoretical study we have investigated the electronic structure of epitaxial LaVO$_3$ thin film at room temperature and also below the structural and magnetic transition temperatures using resonant photoemission spectroscopy (RPES) and density functional theory (DFT) calculations. It is noticed that in the monoclinic phase of the LVO film, besides the dominant 3$d^1$ state, a charge transfer (CT) screened 3$d^2$$\underline{L}$ state also appears the top of the valence band revealing enhanced CT character in contrast to Mott-Hubbard character in the orthorhombic phase. The computed density of states also confirm a stronger V-3$d$ and O-2$p$ hybridization at the valence and conduction band edges in M-LVO as compared to O-LVO. Our calculations show contrasting nature of band gap between the orthorhombic and monoclinic phases. It is observed that a change in crystal symmetry of LVO leads to drastic changes in the electronic structure close to the Fermi level (E$_F$), which can impact the transport properties, while retaining the core electronic structure intact.

**Methodology:**

LaVO$_3$ thin film was grown on (001) oriented single-crystal LaAlO$_3$ (LAO) substrates using the pulsed laser deposition (PLD) technique as describe elsewhere [15]. In order to study the lattice mismatch induced modifications on electronic properties of LVO, the choice of substrate is very critical. The chosen substrate should have large mismatch with LAO and it should not show any structural phase transition in the measured range of temperature, which makes LAO as the



preferred choice [30]. Structural characterization of the grown film was done using θ-2θ X-ray diffraction that confirms the single phase of LVO with an excellent crystallinity. In-plane φ scan further confirms the cube on cube epitaxial growth of LVO on cubic LAO substrate. Importantly, LVO film grown over the LAO is under in-plane compressive strain owing to the lattice mismatch between the LAO and LVO, resultantly the out-of plane lattice parameter is elongated [15]. Ultraviolet photoemission spectroscopy measurements were performed to record the valence band (VB) of LVO film at 140 K and 300 K using an Omicron energy analyzer (EA-125, Germany). VB spectra were recorded at different photon energy values in the range of 30-80 eV at angle integrated photo emission spectroscopy (AIPES) beamline on Indus-1 synchrotron source at RRCAT, Indore, India. The base pressure in the experimental chamber during measurements was of the order of $10^{-10}$ Torr. Prior to the photo emission measurements, the surface of the LVO thin film was cleaned *in situ* using low energy Ar$^+$ ions. The cleanness of the sample surface was checked by the absence of C 1*s* feature and an extra features at higher binding energy feature in O 1*s* core-level spectra, which arises from the contamination or degradation of the sample surface. Although, the surface degradation processes are much slower at low temperature. For calibration of binding energies, Au foil was kept in electrical contact with the sample and the Fermi level ($E_F$) was aligned using the valance band spectra of Au foil. The total instrumental resolution [FWHM: full width at half maximum] was estimated to be ~300 meV at $h\nu$ = 52 eV. The background of the VB spectra was corrected using Shirley method.

Electronic structures of LaVO$_3$ under monoclinic (M) and orthorhombic (O) symmetries were calculated within the framework of density functional theory (DFT) using generalized gradient approximation (GGA) for exchange correlation. The valence *d*-electrons of the transition metal were treated explicitly using on-site coulomb repulsion term *U*, by employing the simplified rotationally invariant approach introduced by Dudarev et. Al [31]. We have used U = 3 eV and J = 0.6 eV in the calculation, which is known to better describe the electronic structure of LVO [10,32,33]. Spin-polarization is employed in all the calculations. Our unit cell lattice parameters for monoclinic and orthorhombic systems were respectively, a = 5.59 Å b=7.75 Å, c=5.56 Å with β = 90.12° and a = 5.558 Å, b=7.834 Å c=5.549 Å [19]. The unit cell was kept fixed and the ions were allowed to relax until the Hellman-Feynman forces were less than 0.005 eV/Å and the energy difference between the successive ionic relaxation steps was less than $10^{-4}$ eV. The



relaxed unit cell was used for calculating the electronic band structure and density of states (DoS). All calculations were performed using a 6 x 4 x 6 k-point set.

## Results and Discussions:

**Low temperature valence band spectrum:**

Fig. 1(a) shows the valence band spectrum (VBS) of epitaxial LVO thin film recorded at photon energy ($h\nu$) 52 eV at 130 K temperature (LT). The measured VBS was fit with six Gaussians, which could effectively reproduce the major features of the spectrum as shown in the Fig. 1 (a). The first broad feature centered at binding energy ($E_B$) 1.5 eV marked as A, is attributed to the dominant V 3$d$ character [15,34]. Broadness of the V 3$d$ band was discussed by Egdell *et al.* [35] in terms of strong electron-phonon interaction, related to the strong polarization of the lattice by a valence electron. Besides, in strongly correlated systems, strong electron correlation effect causes an additional electron-hole excitation on the created hole, which accompanies the photoemission process and consequently broadens the $d$ band structure [36]. The higher binding energy region (from 3 to 9 eV) is mostly dominated by O 2$p$ characters and is fitted with peaks marked as B, C, D and E centered at $E_B$ = 4.7eV, 5.9 eV, 6.9 eV and 8.4 eV respectively. The probable origin of these features are debated, although we have assigned them on the basis of previous experimental and theoretical observations at RT [15,16,34], which are further confirmed from our theoretical calculations as discussed later in the manuscript. The feature B (4.7 eV) is related to the non-bonding character of O 2$p$ band [15,34], the feature C around 5.9 eV is related to the Mott-Hubbard screened 3$d^2$$\underline{D}$ ($\underline{D}$ denotes a hole in the nearest-neighbor V 3$d$ state) configuration [16] and feature D (~ 6.9 eV) is attributed to mixed O 2$p$ and V 3$d$ in nature [34]. The La 5$d$ and 4$f$ bands mostly appear in the conduction band of LVO, so feature E could be related to the La 4$d$/La 5$p$-O 2$p$ hybridized states. Relatively a weak feature is observed at around 10.3 eV assigned as feature F, which is related to the satellite structure of vanadium [37] as observed by Smith and Henrich, in the angle integrated UPS spectra of cleaved $V_2O_3$.

**Temperature dependent VBS:**

The effect of temperature on VBS can be seen in the Fig 1 (b), where we compare the VBS recorded at 130 K with the spectra recorded at 300 K. At 130 K, though the basic features in the valence band do not change significantly and the spectra appear to be similar to that at 300 K, a clear enhancement in DOS weight can be seen in the higher binding energy region (5 to 9 eV). With respect to 300 K, the intensity of the features C and D increases considerably whereas, the intensity of the feature E decreases at 130 K. Thus, it appears that on lowering the temperature, a spectral weight transfer takes place from the mixed O 2$p$-La 4$d$/La 5$p$ state to the hybridized O 2$p$-V 3$d$ states. It is important to mention here that the observed changes do not arise due to temperature dependent surface degradation since in that scenario, one would expect increase in the O 2$p$ feature in this region at high temperature [38]. It should be noted that at room temperature in orthorhombic structure of LVO, La cations are surrounded by 12 oxygen atoms and the La-O distances vary between 2.422 Å and 3.271 Å. However, below first order structural transition (141 K) of LVO, the La-O distance increases and the distortion is so large that not all oxygen neighbors are considered as a first-nearest to La cation [19]. Such changes in the unit cell will cause changes in the spectral intensity features around the hybridization region (O 2$p$-La 4$d$/5$p$ and O 2$p$ –V 3$d$) of the VBS, as is observed. VBS obtained at different photon energy values also reflect the similar kind of spectral intensity modulation with the temperature in the higher binding energy region (features C, D and E). Absence of spectral intensity at the Fermi energy ($E_F$) level at both the temperature values highlights the insulating nature of the LVO film at below and above its structural and magnetic transition temperature, as shown in the inset of Fig. 1 (b).

The low energy spectral features of VB are very crucial in establishing the correlation among structural and electronic properties, thus it is essential to look closely the temperature evolution of spectral structure near $E_F$. For this we show the spectra in Fig. 1 (c) after subtracting the O 2$p$ contributions (solid line in the inset of Fig. 1 (b)) appearing at binding energies higher than 3.5 eV) following the procedures in Refs. [10,39]. The spectral feature centered at about 1.5 eV below $E_F$, normally termed as the incoherent feature, is the spectral signature of the lower Hubbard band (LHB) and corresponds to the electron states essentially localized due to electron correlation [34]. A clear spectral intensity modification is also observed around this energy region of the VB with temperature. It is evident from the spectrum recorded at $h\nu$ =52 eV that on lowering temperature, a minute gain in spectral intensity around 2.3 eV is observed, while the spectral intensity decreases at 1.5 eV. This observation would indicate a transfer of spectral weight from 1.5 eV to 2.3 eV binding energy of the VB at lower temperature. However, an unexpected scenario is observed with the increase or decrease of incident photon energy. For the spectrum recorded with $h\nu$ = 40 eV, a gain in spectral intensity around 2.3 eV binding energy is observed on lowering temperature without any modification at 1.5 eV. Whereas for the spectrum recorded with $h\nu$ = 68 eV, the variation in spectral feature at 2.3 eV is not noticeable on lowering the temperature, though the spectral intensity is observed to decrease at 1.5 eV. These observations clearly suggest that there is no spectral weight transfer from 1.5 eV to 2.3 eV binding energy as was concluded from temperature dependent VBS recorded at $h\nu$ = 52 eV, rather it would be an emergence of new states around 2.3 eV at lower temperature.

We argue here that the emergence of this new feature at lower temperature is related to the structural transition of LVO from RT orthorhombic to low temperature monoclinic phase. As the distortion of the $VO_6$ octahedron increases in monoclinic LVO unit cell compared to RT orthorhombic cell, it will lead to the higher overlap between V 3$d$ and O 2$p$ orbitals [19]. Consequently, O 2$p$ derived states appear along with the V 3$d$ states near the $E_F$ on lowering the temperature. It is known that the photoionization cross section of the O 2$p$ related states enhances as the photon energy is lowered, for example from 52 to 40 eV [40]. Thus the new states distinctly observed in the monoclinic phase at lower photon energy, where the emission cross section of O 2$p$ states is higher, endorse the advent of O 2$p$ states. This is further exemplified by the enhanced intensity of feature A upon further lowering the incident radiation ($h\nu$ = 32 eV) as shown in the inset of Fig. 1 (c), which is not observed at 300 K.

**Angle dependent valence band spectrum:**

It is noted here that the appearance of new state at around 2.3 eV at lower temperature can also be attributed to the surface electronic state, mainly originating due to the surface degradation, albeit the features related to surface degradation mostly appears around the higher binding energy region [38,41]. Apart from degradation, surface states also arise from an enhanced correlation effect near the surface region due to the reduced dimensionality at the surface or the subtle changes of surface geometry [38]. Moreover, the crystal symmetry at the surface is expected to be lower compared to the octahedral field in the bulk, leading to a local $D_{4h}$ symmetry by the means of surface reconstruction [42]. To discard the possibility of surface state attributions to the feature at 2.3 eV, we recorded the angle dependent VBS at 300 K, which can be employed to distinguish the surface electronic structure from the bulk electronic structure [42,43]. We performed the angle dependent photoemission measurements at fixed incident radiation ($h\nu$ = 56 eV). Fig. 2 (a) shows the strong angular dependence of the relative intensities of VBS of LVO/LAO film in the higher binding energy regime (6-10 eV), which are mostly dominated by O 2$p$ states. States related to the surface degradation also generally appear around this region of VB [38,41]. Since the probing depth decreases with the decrease in the incident angle, the spectral intensity monotonically increases around this region (6 to 10 eV). This is consistent with the notion that at lower incident angle, the surface states contribution should enhance. In contrast, as shown in the inset of Fig. 2 (a), the region (0 - 2.9 eV), where the spectral intensity is mostly dominated by the V 3$d$ emission, reveals highest intensity for $\varphi$= 45˚, which gradually reduces upon decreasing the incident angle as also seen even after O 2$p$ subtraction in Fig. 2 (b) [42]. Such variation in spectral intensity of the VB strongly suggests that the features in the low binding energy range are not due to the surface states, rather indicates appearance of a new feature related to the structural transition as discussed above.

**Resonant photoemission spectroscopy measurement:**

To better understand the spectral features and observed variation in the VBS we have performed the RPES measurements at 130 K. RPES is a tool to distinguish 3$d$ states form a complex hybridized states and subsequently we have also been able to figure out the O 2$p$ dominant sates [44]. For RPES study, VBS of the epitaxial LVO film were recorded for different photon energy swept through V 3$p$ →3$d$ excitation. The energy distribution curves (EDCs) of the epitaxial LVO film for the photon energy varying from 30 eV to 75 eV show the variation in spectral features (not shown here). The photon energy dependence of the spectral intensity is revealed more clearly in the constant initial state (CIS) spectra. It is important to mention that the RPES and CIS spectra were normalized to the intensity of O 2$p$ non-bonding state at binding energy 4.7 eV [15]. With the variation of incident photon energy, the CIS of feature A shows a strong resonance enhancement with maximum around 52 eV preceded by a dip at around 42 eV, as shown in the Fig. 3 (a). The resonance of feature A can be understood by the quantum-mechanical interference between the direct photoemission from 3$d$ state. V: $3p^6 3d^2$ ($t_{2g}^2$) + $h\nu$ → V: $3p^6 3d^1$ ($t_{2g}^1$) + e$^-$ and the intra-atomic excitation process by 3$p$ state followed by super Coster-Kronig decay V: $3p^6 3d^2$ + $h\nu$ → $[3p^5 3d^3]^*$ → V: $3p^6\ 3d^1$ + e$^-$ processes. Two excitation processes transform a certain initial state to the same final state and consequently the V 3$d$ photoelectron yield increases dramatically and exhibits resonance. Hence, the resonance enhancement of the feature A at 52 eV photon energy confirms the dominant V 3$d$ character at the 1.5 eV binding energy of the VB. Besides, a rise in intensity is also observed with decrease in photon energy below 42 eV. In fact the O 2$p$ photoionization cross-sections increases with the decrease of photon energy [40], which suggests the presence of considerable amount of O 2$p$ states in the form of hybridized with V 3$d$ at the 1.5 eV binding energy. Such enhancement of CIS intensity at lower photon energy was conspicuously absent at room temperature [15]. This observation indicates the enhanced contribution of O 2$p$ states in the monoclinic phase in the form of its hybridization with the V 3$d$ states as well as appearance of new feature at 2.3 eV as discussed above. To further understand the CIS profile of feature A in detail, we have performed the Fano line fitting and compared it with the RT CIS and discuss later in the manuscript.

The CIS profile for the feature D in the O 2$p$ dominant region of VB also reveals resonance enhancement at around 52 eV photon energy as shown in the Fig. 3 (b). Generally, oxygen derived states will not show a resonant enhancement in the intensity around this photon energy interval. However, resonance in the ligand emission, upon sweeping the photon energy through the metal 3$p$ to 3$d$ transition, is attributed to the metal $d$ state hybridization with the ligand states [45]. Thus observed resonance in CIS of feature D ($E_B$ = 6.9 eV) suggests the presence of an appreciable amount of V 3$d$ character as well. Such resonant enhancement in O 2$p$ band was also observed in YVO$_3$, CaVO$_3$ and other vanadium oxides [45,46] that was attributed to O 2$p$ and V 3$d$ hybridization. CIS spectra of other features like B ($E_B$ = 4.7 eV), C ($E_B$ = 5.9 eV), E ($E_B$ = 8.4 eV) and F ($E_B$ = 10.4 eV) do not reveal any such resonance enhancement in intensity in the photon energy range of 40 to 60 eV, as shown in Fig. 3 (c), pointing out towards their O 2$p$ character [15].

The shape of the resonance profiles obtained from RPES are used to distinguished between the 3$d^{n-1}$ and 3$d^n\underline{L}$ (where $\underline{L}$ denotes a ligand hole) final states associated with the particular feature in VBS [47-50]. As has been demonstrated for heavy transition-metal (TM) compounds [48,51], the CIS spectra for 3$d^{n-1}$ final states usually show only Fano-type resonance peak without a remarkable anti-resonance dip near the TM 3$p$ →3$d$ threshold while for 3$d^n\underline{L}$ final states an anti-resonance dip on lower photon energy side of shallow peak is accentuated. However, the strong screening of the 3$d$ hole by the charge transfer (CT) from anion atoms (3$d^n\underline{L}$) has been clearly found in the photoemission spectra of heavy-TM compounds, owing to the smaller value of CT energy [48,51,52]. On the other hand, such screening effect is usually less evident for light-TM compounds due to the less covalent character or the higher CT energy that might prevent such CT screening for light-TM compounds [45,52]. In our resonance experiments of LVO thin film at 130 K, the feature A (1.5 eV) and feature D (6.9 eV) show resonance enhancement with the photon energy, though a clear difference is present in their resonance profile. Resonance of the feature D shows a broad and simple enhancement in intensity without any anti-resonance dip, even though the final state of the photoemission from this O $2p$ band is $3d^n\underline{L}$ final state. Such intriguing behavior is rather common for the vanadium oxides [15,45,46]. In contrast to much narrow and asymmetric shapes observed for heavy-TM compounds, the broad resonance profile extended over a large energy range of light-TM compounds is attributed to the shakeup excitation that essentially acts as a loss accompanying the $3p \rightarrow 3d$ optical absorption [53]. On the other hand, the resonance profile of feature A reveals an anti-resonance dip around at 42 eV photon energy followed by a sharp resonance, suggesting that $3d^2\underline{L}$ ($3d^n\underline{L}$) sates overlap with the dominant $3d^1$ ($3d^{n-1}$) states at 1.5 eV binding energy [49,50]. Therefore, the top of the VB of LVO thin film at 130 K reveals a considerable hybridization between V $3d$ and O $2p$, different from the room temperature VB, where only dominant V $3d$ band with $3d^1$ final state was present, as shown in the Fig. 3 (d) [15]. One of the most striking differences between the monoclinic and orthorhombic phase of bulk LVO is the number of equivalent sets of V sites. In orthorhombic phase, all V sites are equivalent, while in the monoclinic phase, two sets of in-equivalent V sites form, which are ordered along the *b* axis of the monoclinic cell [19]. Besides, in monoclinic structure of LVO, the distortion as well as the tilting of $VO_6$ octahedra increases that might enhance the covalent charter of V, which could be the possible reason for the modification in VBS and emergence of hybridized oxygen states at lower binding energy regime in the VBS at 130 K.

To further understand the LT and RT resonance profiles of feature A, we fit the CIS with characteristics Fano-line profile [54]. CIS spectra of feature A is fitted with the Fano line shape profile

$$I(hv) = I_a \frac{(q+\varepsilon)^2}{1+\varepsilon^2} + I_b$$

With, $\varepsilon = \frac{h\nu - E_m}{\tau}$, where $E_m$ is the resonance energy, $\tau$ is the line width depending on the decay rate of autoionization process, and $I_a$ and $I_b$ are the non-resonant background intensity for the transition to continuum state that interacts or does not interact, respectively, with discrete autoionization state. $q$ is the asymmetry parameter defined as $\frac{\langle v|r|i\rangle}{\langle v|V|f\rangle\langle f|r|i\rangle}$. It is the ratio of transition from the initial state ($|i\rangle$) to the discrete state ($|v\rangle$) and the initial state ($|i\rangle$) to continuum ($|f\rangle$) interacting with the discrete state ($V$ is the Coulomb interaction) and r is the transition operator. The Fano parameter $q$ value depends on the shape of resonance profile and represents the discrete/continuum mixing strength, i.e., the coupling strength. From the Fano line profile fitting, as shown in the Fig. 3 (d), we observed $q$ value about 1.2 and 2.3 for 130 K and 300 K respectively. The minute discrepancies in Fano line fitting arise mostly due to the broadness and shifting of the resonance maximum from the $3p \rightarrow 3d$ threshold by about 10 eV due to exchange interaction between the $3p$ hole and n+1 $3d$ electrons [55]. The exchange interaction is so large for light-TM compounds owing to the extended nature of the $3d$ wave function, that it causes the resonance peak far above the threshold by raising the optically favored $3p^5 3d^3$ multiplets and consequently a delay of about 10-15 eV in onset of resonance [55].

It has been demonstrated, if the CIS plot consists of only Fano-type resonance, the $q$ value becomes large which infers that the coupling strength between the continuum state ($|f\rangle$) and the discrete state ($|v\rangle$) is very weak. Whereas, presence of anti-resonance in CIS profile followed by the Fano resonance leads to a lower value of $q$ and hence strong discrete/continuum coupling strength [56]. Therefore, the obtained higher $q$ value and the absence of remarkable anti-resonance dip in the RT CIS of feature A (1.5 eV) confirm the dominant V $3d$ character with V $3d^1$ final state [57]. In contrary, comparatively lower $q$ value and the noticeable anti-resonance dip associated with the LT CIS of feature A (1.5 eV) indicate the inclusion of considerable V $3d$ - O $2p$ hybridized state with V$3d^2\underline{L}$ final state into the dominant V $3d$ character. So at 300 K, the top of the VB of LVO thin film is dominated by V $3d$ character with $3d^1$ final state, hence it was ascribed as a lower Hubbard band (LHB). Combined spectra of occupied and unoccupied states infers the $d$-$d$ type low energy charge fluctuation which confirms the Mott-Hubbard type of insulating character of LVO thin film at room temperature [15]. However, at 130 K the top of the VB comprises of considerable amount of $3d^2\underline{L}$ state mixed with the dominant $3d^1$ final state suggesting enhanced CT characteristics in monoclinic LVO [58].

**Discussion:**

In order to validate our experimental observations and gain microscopic insights on the difference in electronic structures of monoclinic (M-LVO) and orthorhombic (O-LVO) LVOs, density functional theory calculations were carried out. Although experimentally the O-LVO is known to be paramagnetic at room temperature, non-magnetic calculations lead to a metallic behaviour and hence a magnetic solution is found to be necessary to obtain the correct local $3d$ electronic structure or the insulating character consistent with experiments [59]. Since the bulk LVO is known to stabilize in $C$-AFM state, the same magnetic state is used in our calculation for both the cases. There are four formula units (f.u.) in a LVO unit cell, accordingly the vanadium ions are numbered V1 through V4 as shown in the Fig. 4 (a), as viewed along the {101} direction. The M-LVO structure is energetically more favourable between the two structures with an energy gap of 15.67 meV/f.u. The electronic band structures of these two unit cells are very distinct as evident from Fig. 5 (a) and 5 (b). The O-LVO shows a smaller band gap of 0.88 eV compared to the M-LVO counterpart with a gap of 1.27 eV. The calculated band gaps are also comparable to the previous experimental results [15,32]. Interestingly the O-LVO unit cell shows a direct gap at the Brillouin zone centre or the Gamma point. In M-LVO, the valence band (VB) maximum occurs at the R point, however the conduction band (CB) minimum occurs between the R and X points. It is apparent that the nature of the band gap alters from direct in O-LVO to indirect in M-LVO. The change in crystal symmetry from monoclinic to orthorhombic lifts the near degeneracy of bands around the Gamma point both in valance and conduction bands leading to a direct gap in O-LVO as well as to the reduction of band gap. The direct nature of band gap in O-LVO as discussed above can be very advantageous for photovoltaic applications over the monoclinic phase for efficient photon absorption, without involvement of phonon for momentum conservation [60]. The VB and CB show large dispersions in O-LVO compared to the M-LVO cell. The CB of M-LVO is more or less flat and isolated in energy space suggesting localized nature of the band. Further analysis shows that the top of VB and bottom of CB are mainly of vanadium $3d$-orbital character. We also notice that the near degeneracy of bands around the Gamma point is completely lifted in case of O-LVO structure, leading to large band dispersions. This is corroborated to well spread out density of states in the O-LVO unit cell (shown alongside the band dispersions), suggesting a larger band width of metal

*d*-orbitals. For the M-LVO system we find relatively narrow VB and extremely sharp CB. Our calculations show that the electronic states that are close to the $E_F$ undergo significant changes.

In Fig. 5 (c) and 5 (d) we show the calculated projected-DOS (PDOS) of V 3*d*, O 2*p* and La 4*d*, 5*p*, 5*d*, 4*f*. The DOS around the Fermi level is zoomed and shown in the inset. The calculated DOS of LVO below the $E_F$ is consistent with the experimentally observed VBS. In both the cases the dominant V 3*d* bands are split across the $E_F$ due to correlation effects (inset in the figure 5(c) and (d)), suggesting a Mott-Hubbard type insulating character [15,34]. However, there exists a minor contribution from O 2*p* orbitals due to V 3*d* – O 2*p* hybridization, which is more pronounced in M-LVO compared to the O-LVO structure, in agreement with our VBS results. An increase in the O 2*p* contribution in the VB edge indicates an enhanced CT character in case of M-LVO, as compared to the O-LVO as shown schematically in Fig. 5 (e) from our combined experimental and theoretical observations. This enhancement of metal-ligand hybridization could play a crucial role in deciding the nature of magnetic exchange as well as transport properties of LVO. Unlike in Mott-Hubbard insulators where the strong coulomb interaction between the *d*-electrons decides the electronic structure and properties, the properties of a CT insulator are controlled by the CT gap Δ, which can provide an additional degree of freedom to optimize material properties [52]. These bands are seen within 1 eV below and just over 1 eV above $E_F$ respectively and qualitatively corresponds to the feature A of our VBS. A considerable difference in the spread of DOS and presence of O-2*p* PDOS across the $E_F$ is more evident in M-LVO structure. Our RPES analysis at 130 K also suggests the CT screened $3d^2\underline{L}$ states are mixed with the dominant $3d^1$ final state for the spectral feature appeared at 1.5 eV below the $E_F$. Major contribution of O 2*p* bands to the total DOS extend from 2 to 6 eV below $E_F$. The La-4*d* and La-5*p* contributions appear between 4 to 5 eV below $E_F$, this combined with the O-2*p* states correspond to feature E or the La-5*p* and O-2*p* hybridized band. In between 2 and 4 eV, we notice DOS dominated by only O-2*p* contributions similar to the non-bonding character in feature B. Below this we also see weights for V-3*d* - O-2*p* hybridization of feature D. The La 4*f* and 5*d* bands appear 3.0 eV above $E_F$. Our calculated DOS can thus be mapped one-to-one with the experimental VBS spectrum discussed in the previous section.

Analysis of unit cell structure of these two structures provide insights into the enhancement of V 3*d* – O 2*p* hybridization in M-LVO compared to O-LVO as shown schematically in Fig. 4 (b). In O-LVO among six V-O bonds in a $VO_6$ cage there are three distinct bond lengths, $d_{V-O}$ = 1.99 Å, 2.01 Å and 2.04 Å, of which the shortest and longest bonds are along the *ac*-plane while the $d_{V-O}$ = 2.01 Å is along the *b*-axis. In M-LVO, the corresponding values are $d_{V-O}$ = 1.99 Å, 1.99 Å and 2.06 Å. In this structure, we notice a slight contraction of the bond along the *b*-axis and correspondingly an elongation of one of the bonds along *ac*-plane, consistent with the shorter *b*-axis lattice parameter. The bond angles $\theta_{V-O-V}$ in M-LVO is about 154.2° along the *b*-axis as well as in the *ac*-plane. However, in the O-LVO, while the in plane $\theta_{V-O-V}$ remains nearly unchanged at 154.1°, the *b*-axis $\theta_{V-O-V}$ is reduced to 152.9°. Thus a marginal decrease in one of the bond lengths and the proximity of $\theta_{V-O-V}$ bond angles to 180° can provide M-LVO an edge over O-LVO in terms of stronger V 3*d* – O 2*p* hybridization [41].

Such modifications in unit cell of M-LVO and O-LVO structure can have potential impact on the V3*d* orbital DOS [33]. To visualize the effect of such distortion in the form of tilting and rotation of oxygen octahedron on V 3*d* orbital DOS near across $E_F$, the calculated DOS of V-3*d* ions V1 through V4 are shown in Fig. 6 (a, b) for the M-LVO and O-LVO structures. The ions V1 and V4 (V2 and V3) are connected by oxygen ions along the *b*-axis, while V1 and V2 (V3 and V4) are connected along the *a-c* plane as shown in the Fig. 4 (a). Cooperative Jahn-Teller (JT) distortion of $VO_6$ octahedra can lift the degeneracy of $t_{2g}$ orbitals and put the $3d^2$ electrons into two lower energy orbitals among $d_{xy}$, $d_{yz}$ and $d_{zx}$ orbitals [61]. Further, LVO is also known to undergo $GdFeO_3$-type distortion leading to rotation and tilting of the $VO_6$ cage. This results in lowering of the local symmetry from $O_h$ to $D_{4h}$, leading to mixing of $t_{2g}$ and $e_g$ orbitals as well [33]. In our DOS calculations for both M-LVO and O-LVO structures we see nearly equal weights for $d_{x^2-y^2}$ and $d_{3z^2-r^2}$ orbitals compared to the $d_{yz}$ and $d_{zx}$ orbitals in the VB [61]. Moreover, the enhanced lattice distortion in M-LVO structure leads to the splitting between the DOS of $t_{2g}$ and $e_g$ orbitals are evident near $E_F$ in the VB of M-LVO [33]. Besides, in monoclinic structure, we notice that the $d_{yz}$ and $d_{zx}$ orbitals alternatively have large weights when we move from one atomic layer to another along the *b*-axis, thus forming an A-type orbital ordering (OO) [61]. No such OO is evident in the orthorhombic structure. It has also been previously suggested that the orbital fluctuations are quite strong in O-LVO phase of LVO which can be suppressed by forming corresponding OO only in the M phase by both JT and $GdFeO_3$-type distortion [10]. Such observation of orbital ordering can have huge implications on the magnetic properties as well [61], which we relegate to a future communication.

Heterostructures comprising of indirect band gap M-LVO can be used in Mott electronics, where correlated electron phenomena could be incorporated in device geometrics [62]. Earlier, in contrast to the insulating nature of bulk and LVO film grown on LAO, metallic conductivity was observed at the interfaces of LVO/STO films [22,28,29]. Oxygen vacancies in the STO or the charge reconstruction at the interface was believed to be the cause for such enhanced conductivity [29,63]. The dispersion less nature of conduction bands in M-LVO as shown in Fig. 5 (c) means a large effective mass of electron in comparison to the O-LVO, which shows large band dispersion. As the conductivity is related to effective mass by, $\sigma = \frac{ne^2\tau}{m_{eff}}$ and the effective mass is inversely related to the second derivative or the curvature of the energy bands, the O-LVO is expected to have better conductivity than M-LVO. In other words, the conductivity is believed to be strongly suppressed in M-LVO up on lowering the temperature. Our present observation thus throws new light on change in conductivity below the first order structural transition temperature [22,28].

**Conclusion:**

We have probed the low and room temperature occupied electronic states of the epitaxial LaVO$_3$ thin film grown on LaAlO$_3$ substrate. The photon energy dependence of the valence band spectra obtained at room temperature orthorhombic state of LVO confirms the dominant V 3$d$ character with 3$d^{n-1}$ final state, while a CT screened 3$d^n\underline{L}$ final state mixed within the dominant V 3$d$ character for low temperature monoclinic LVO. Calculated DOS for both O-LVO and M- LVO structure within the framework of DFT well describes the experimental valence band spectrum. The obtained bands across E$_F$ has dominant V 3$d$ character and are attributed as the LHB and UHB, confirming a Mott-Hubbard insulating nature with *d-d*-type of low energy charge fluctuation for both structures. The monoclinic structure shows an enhanced CT character than O-LVO due to stronger V 3$d$ - O 2$p$ hybridization due to larger octahedral tilt in M-LVO. The orthorhombic structure shows a direct band gap and is expected to have superior conductivity than the monoclinic structure. Our present results indicate a scenario for LVO in which the electronic properties are strongly coupled with the lattice degree of freedom.

**Acknowledgements:**

Authors acknowledge Mr. A. Wadikar and Mr. Sharad Karwal, for their help during RPES measurements. Science and Engineering Research Board, Department of Science and Technology, India is acknowledged for the financial support under the grant: CRG/2019/001627.

**Figures captions:**

**Figure 1:** (a) Valence band spectrum (VBS) of LaVO$_3$ thin film recorded at 52 eV photon energy at 130 K (LT), inset shows the zoomed view near E$_F$. (b) Comparison of LT and RT VBS of LVO thin film recorded at different photon energies (40, 52 and 68 eV), top inset shows the zoomed view of the LT and RT spectral feature near E$_F$ and bottom inset shows the expanded tailing of O 2$p$ states. (c) After O 2$p$ subtraction: comparison between LT and RT VBS near the E$_F$ recorded at different incident photon energy.

**Figure 2:** (a) Schematics of the angle dependent photoemission plane. (b) VBS of the LVO thin film recorded at different incident angle. (c) The angle dependent VBS near E$_F$ after O 2$p$ subtraction.

**Figure 3:** (a) and (b) CIS plot of feature A (1.5 eV) and feature D (6.9 eV) obtained at 130 K. (c) CIS plot of feature B (4.7 eV), C (5.9 eV), E (8.4 eV) and F (10.4 eV) recorded at 130 K. (d) CIS plot of feature A (1.5 eV) obtained at 130 K and 300 K along with the Fano-line shape fitting.

**Figure 4:** (a) Schematics of monoclinic and orthorhombic unit cell of LVO along with a *C*-type antiferromagnetic spin configuration, as viewed along the {101} direction. V-O-V bond angles of relaxed structures are also shown (b) V-O bond lengths of relaxed monoclinic and orthorhombic structures.

**Figure 5:** (a) and (b) Band structure, (c)-(d) projected density of states (PDOS) of *C* antiferromagnetic (C-AFM) state of M-LVO and O-LVO (e) Schematic of the electronic structure of Mott-Hubbard type insulating LVO for both monoclinic and orthorhombic structures.

**Figure 6:** The projected density of states (PDOS) of V 3$d$ orbitals with *C*-antiferromagnetic (AFM) of LVO for (a) monoclinic and (b) orthorhombic structures.

**Figure 1:**

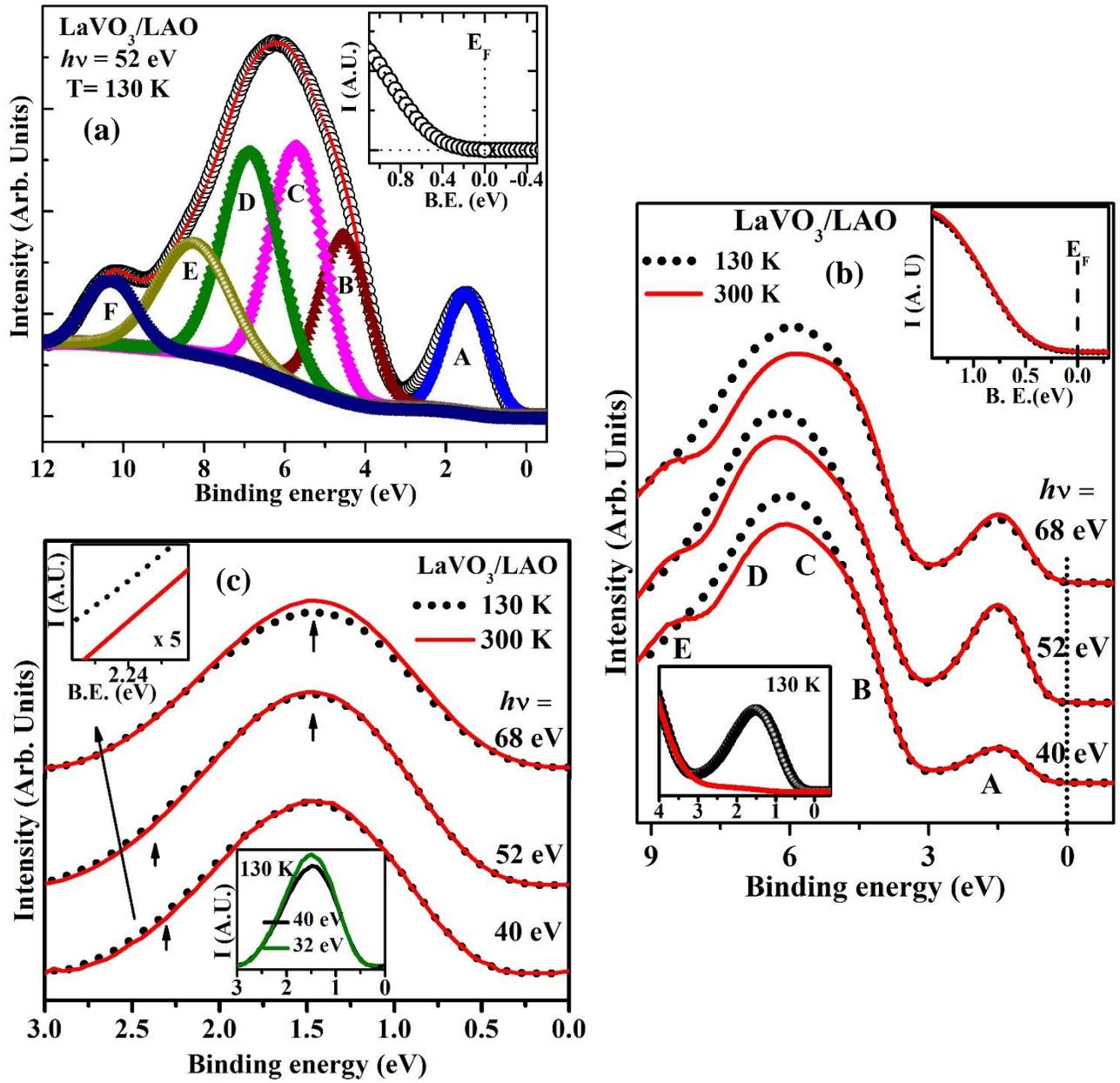

**Figure 2:**

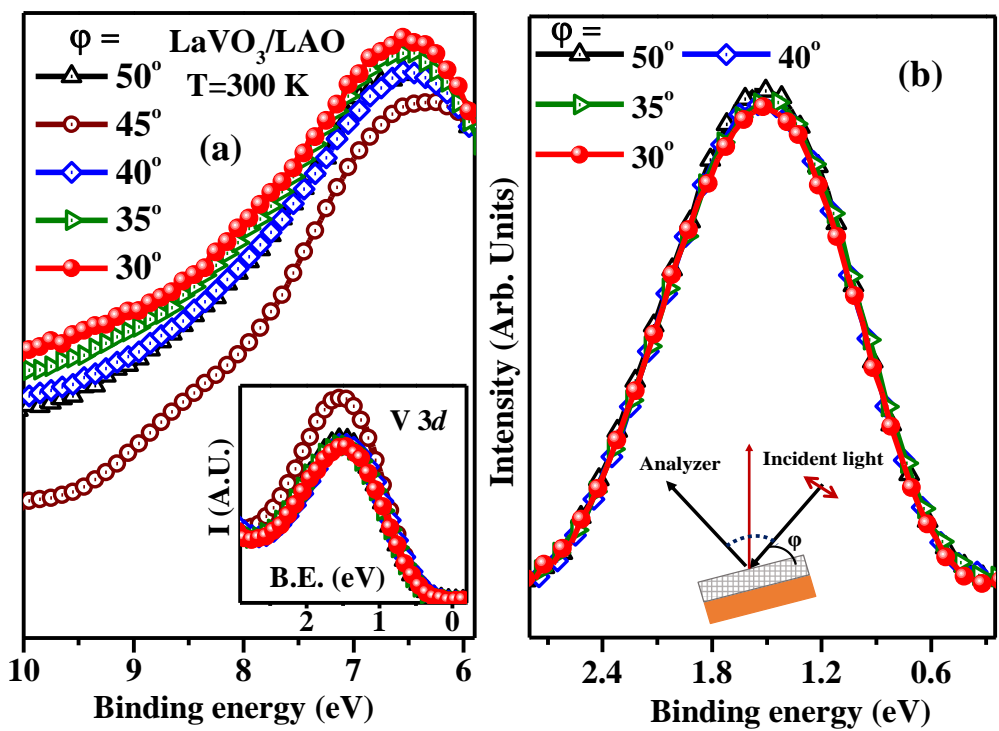

**Figure 3:**

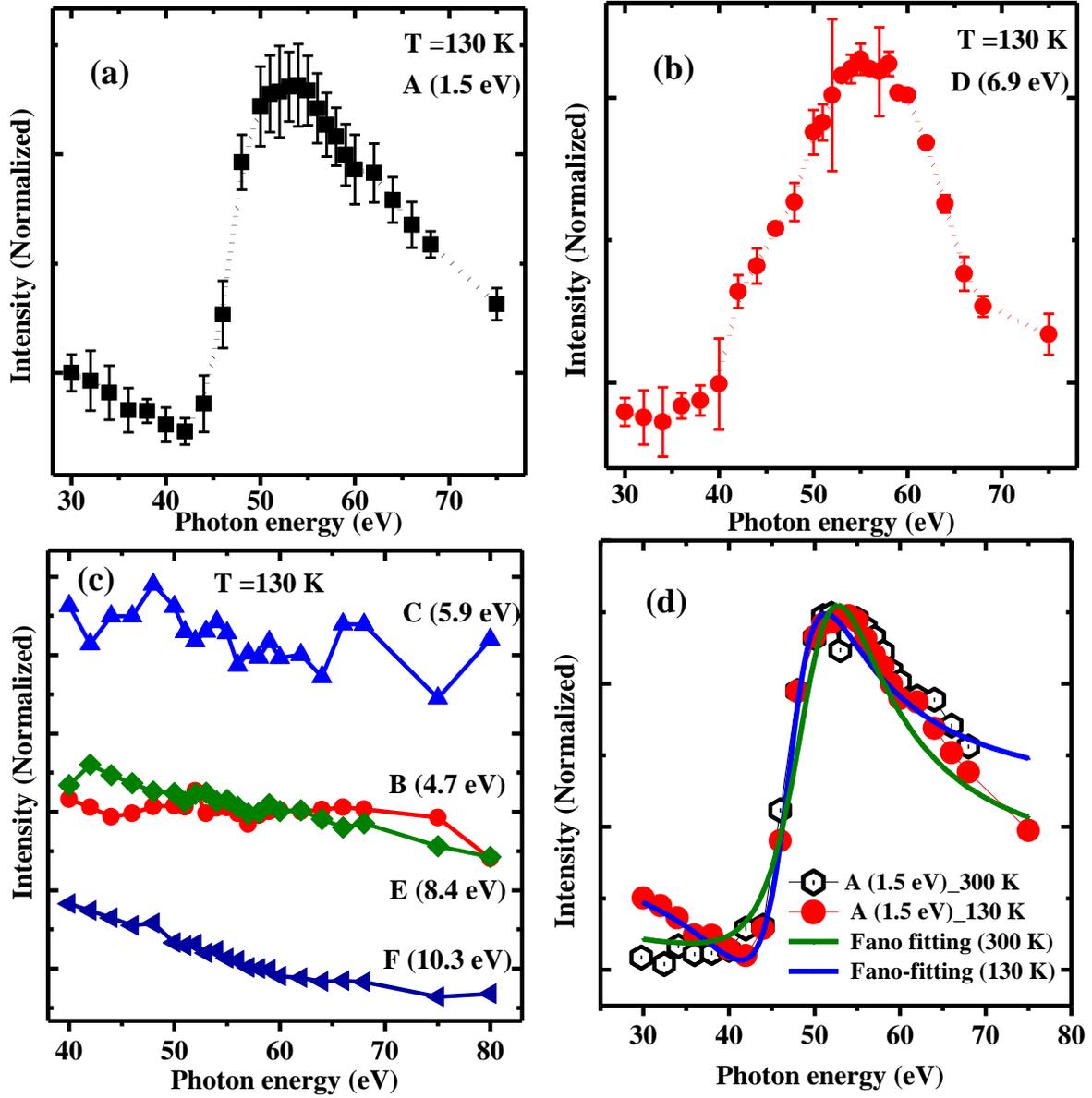

**Figure: 4**

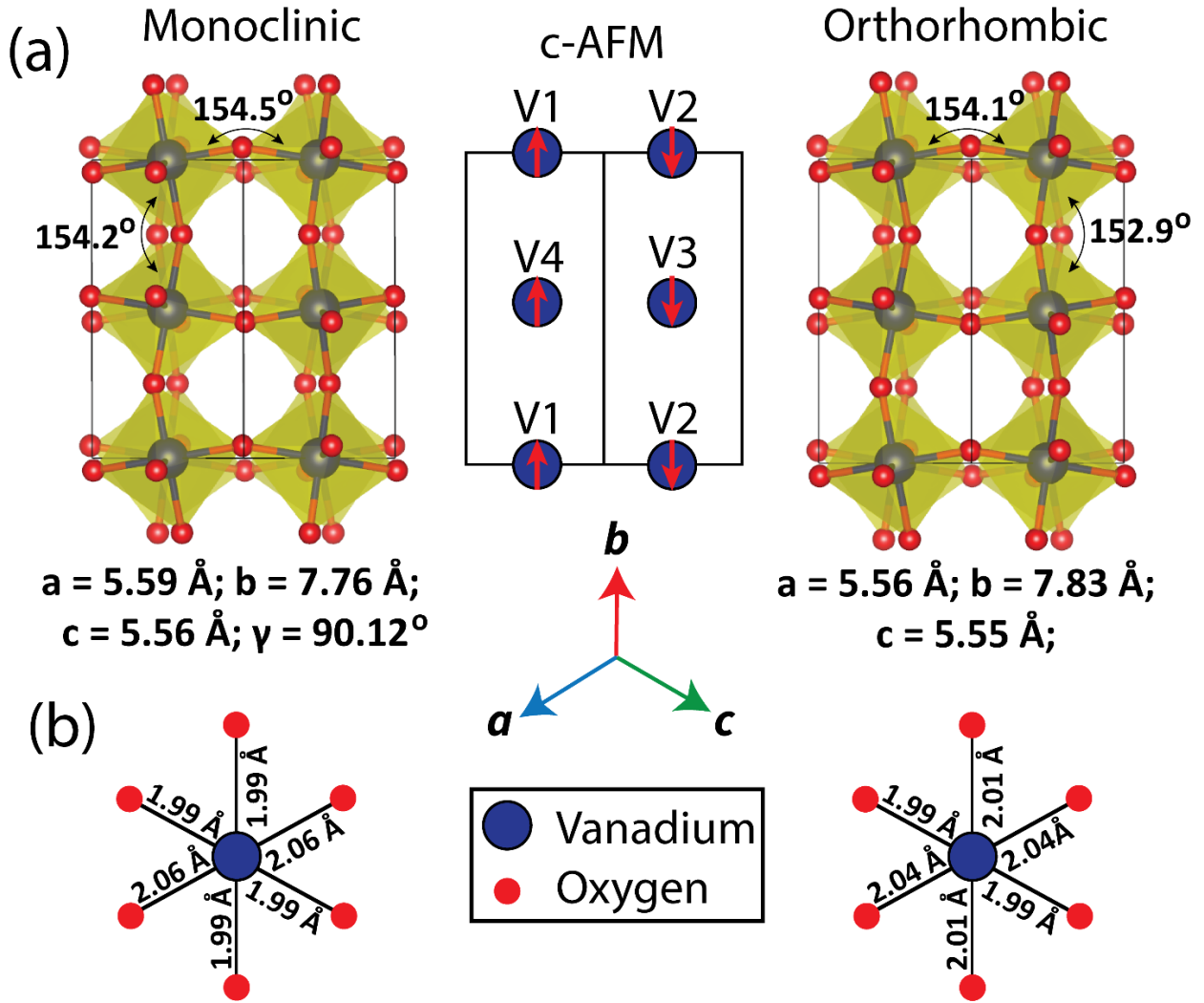

**Figure 5:**

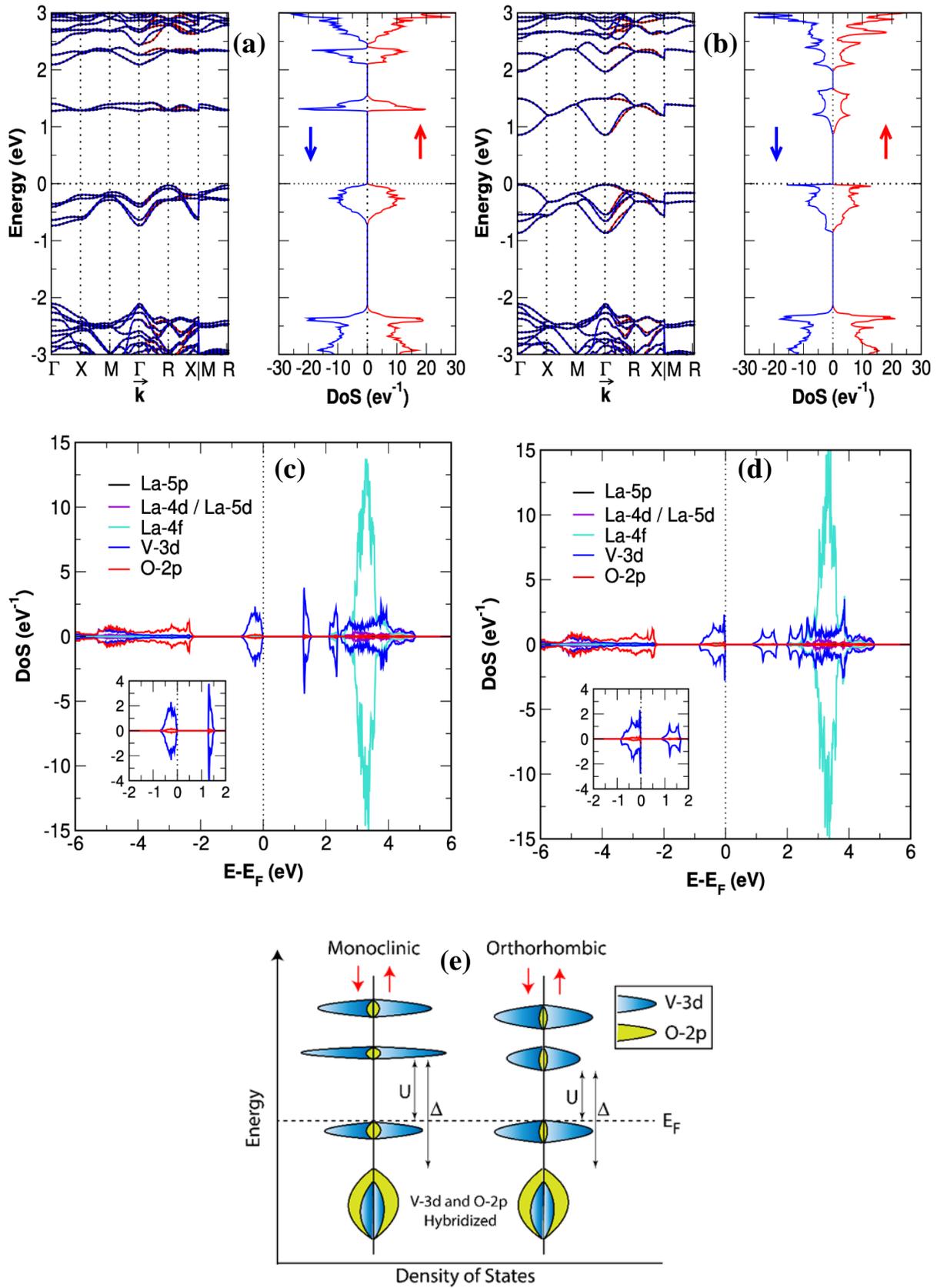

**Figure 6:**

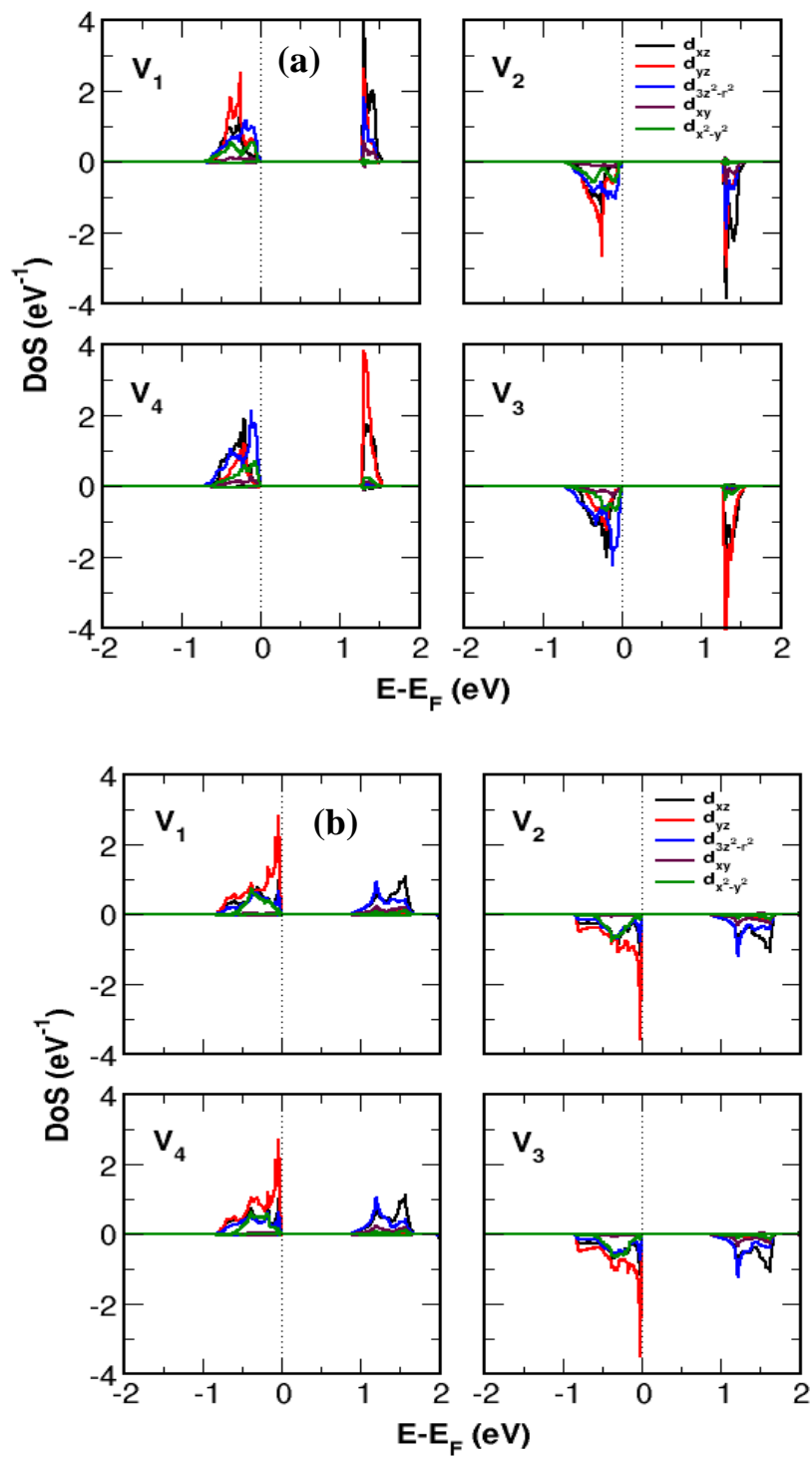